\documentstyle[12pt]{article}

\begin{document}
\title{\Large\bf Wave Function for the Reissner-Nordstr\"om 
Black-Hole}
\author{{\large\sf 
 P.V. Moniz}\thanks{e-mail: {\sf 
prlvm10@amtp.cam.ac.uk; 
paul.vargasmoniz@ukonline.co.uk}}~\thanks{WWW-site: {\sf 
http://www.damtp.cam.ac.uk/user/prlvm10}}\\ 
Department of Applied Mathematics and Theoretical Physics\\ 
University of Cambridge\\ Silver Street, Cambridge, CB3 9EW 
\\ United Kingdom}
\date{{\em DAMTP R/97/23}}
\maketitle

\begin{abstract}
We study the quantum behaviour of 
Reissner-Nordstr\"om (RN) black-holes 
interacting with a complex scalar field. 
A Maxwell field is also present. 
Our analysis is based on M. Pollock's \cite{pol} method 
and is characterized  
by solving a Wheeler-DeWitt 
equation in the proximity of an apparent horizon of the RN space-time. 
Subsequently, we obtain 
a wave-function $\Psi_{{\rm RN}}[M, Q]$ representing 
the RN black-hole
when its 
charge,  $\mid Q \mid$,  is small in 
comparison  with 
its mass, $M$. We then compare  
quantum-mechanically the cases of $(i)$ $Q = 0$ and 
$(ii)$ $M \geq  \mid Q \mid \neq 0 $. A special 
emphasis is given to the evolution of 
the mass-charge rate  affected  by  Hawking radiation.
\end{abstract}

\section {Introduction}

\indent

Recently, there has been a renewed interest 
in the canonical quantization of black-hole 
space-times \cite{pol} - \cite{moniz1}. 
A description of earlier works can be   found 
in 
ref. \cite{ryan} and the general 
purpose of the current  line of   research is to 
provide an adequate framework to study the 
last  stages of gravitational 
collapse \cite{pol}--\cite{moniz1}. 
More precisely, the aim is to obtain a  description 
of quantum black holes that would 
 go beyond a semi-classical 
approximation,  where the background 
metric is treated classicaly \cite{bd}. 
A classical Hamiltonian formulation constitutes an essential step 
in this line and several versions can be found in ref. 
\cite{pol}-\cite{moniz1}, \cite{thomhaji,haji},
with the particularity that several 
matter Lagrangeans are thoroughly   treated in \cite{thomhaji, haji}.
As far as a quantum analysis is concerned, different 
perspectives have also been employed: 
$r$-Hamiltonian quantization \cite{cava}, 
quantization on the apparent 
horizon \cite{pol, tomi, odat, odasolo}, reduced 
phase space quantization (solving the constraints at 
classical level and isolating the physical degrees of freedom)
 \cite{kk, louko, makela, 
louko2}, and also  from  Ashtekar 
variables \cite{kastpthim}.

In this letter, we will extend M. Pollock's 
method \cite{pol} (which was itself influenced by  
the prior  work of A. Tomimatsu \cite{tomi} on 
Schwarszchild space-times) to 
 Reissner-Nordstr\"om (RN) black-holes 
and test it in this particular case. 
As a consequence, we will  find a  wave function for the RN 
 black-hole, 
which will have an explicit dependence on 
its  mass $M$ and the charge $Q$.

Our motivations are  twofold. 
On the one hand, a RN black-hole 
with $M = \mid Q\mid $ has 
supersymmetric properties \cite{susy1, susy2} while 
one with $M > |Q|$ does not. 
Hence, maybe a wave function for the RN black hole could 
provide us with crucial insights of how  
black-hole quantum states in N=2 in 
supergravity would look like \cite{moniz1}. On the other hand, 
a wave function  
$\Psi_{{\rm RN}}[M, Q]$ 
for the 
RN black hole 
could also be used 
to 
discern how its 
mass $M$ varies  with respect to a  time 
coordinate and is influenced 
by the presence of a charge $Q$: $\dot {M} = f(M, Q)$. 
Concerning Schwarzschild black-holes, 
an expression for the 
variation  of $M$ with respect to a time coordinate
due to the back reaction of the Hawking 
radiation \cite{hawking} has been determined 
semi-classically (see e.g., \cite{hawking,bd}). This was further   
validated from quantum gravitational models 
\cite{pol, tomi, odat, odasolo}. A similar treatment 
 regarding charged black holes 
has been very recently and independently presented  in 
ref. \cite{odat, odasolo}. This particular topic contains 
significative physical relevance: as a charged
black-hole emits radiation then 
$M\rightarrow |Q|$ and its temperature 
$T\rightarrow 0$. Hence, its mass evaporation process at this point  
could stop. The presence of the charge 
in a relation as $\dot {M} = f(M, Q)$  may be  essential to 
reproduce this physical effect point to others (see, e.g., ref. 
\cite{minf}).

issues of cosmic censorship \cite{cosmcen} and 
Our approach will have  distinct features 
from those presented in ref. \cite{odat, odasolo}. 
More precisely, this letter  is then organized as follows. 
In section 2 we briefly summarize the 
essential elements of  the 
Hamiltonian formulation of charged 
black-holes in the presence of a matter 
Lagrangean, which is  constituted by complex 
scalar fields and a Maxwell (electromagnetic) 
vector field. We then express in section 3
this framework in terms of 
a RN apparent horizon. Basically, we 
will employ a RN-Vaidya metric \cite{RNvad, fayos, hisc} to describe 
the evaporating black-hole. M. Pollock's method 
\cite{pol} is then appropriately  adapted to our case
 and subsequently used\footnote{Quite recently, A. 
Hosoya and in particular I. Oda \cite{odat, odasolo} have 
also independently analysed
Schwarschild and  RN black-holes from a quantum gravitational point of view. 
They employed the  Hamiltonian formulation of ref. \cite{thomhaji, haji} 
but their quantum mechanical analysis followed instead the 
one introduced by A. Tomimatsu \cite{tomi}.} and tested. 
The relevant dynamical 
quantities and constraints are then obtained
We keep the complex scalar fields  and  
the vector field but within specific approximation 
limits\footnote {Namely, when the 
charge $\mid Q \mid$ is small enough so  
that $M + \sqrt {M^{2} - Q^{2}} \simeq 2M 
- \frac {Q^{2}}{2M}$.}. 
We will solve  the constraint
equations and obtain  wave 
functions $\Psi_{{\rm RN}} [Q, M]$, 
which will constitute solutions of the constraints up 
to terms of the order of $Q^{2} / M^{2}$.
This is done  in section 4, where 
we will  compare the cases of 
 $ Q = 0$ and $M \geq  |Q|$. 
We will also draw some comments on 
how does  $M$ change  with respect to a 
time coordinate.
Finally, we present our conclusions in 
section 5.

\section{Review of the Hamiltonian formulation of RN black-hole}

\indent 

As mentioned above, we briefly review here the main elements 
of the Hamiltonian formalism for charged black-holes, following the 
construction introduced in ref. \cite{thomhaji, haji} 
(see also ref. \cite{pol, tomi, odat, odasolo}). 
The 4-dimensional action has the form 
\begin{equation}
S = \int \ d^4 x \sqrt{- \hat g} \left[ {1 \over 16 \pi}  \hat 
R - {1 \over 
4 \pi} \hat g^{\mu\nu} (\hat D_{\mu} \hat \psi)^{\dag} \hat D_{\nu} \hat  
\psi  -  {1 \over 
16 \pi e^2} \hat F_{\mu\nu} \hat F^{\mu\nu} \right],
\label{eq:jap1}
\end{equation}
where the ``overhat'' denotes a 4-dimensional 
variable,  $\hat \psi$ is a complex scalar field, $\hat 
A_{\mu}$ is the electromagnetic potential and 
$\hat F_{\mu\nu}$ the corresponding field strength, 
$e$ is the electric charge,
$\hat g_{\mu\nu}$ is the 4 dimensional space-time 
metric whose Ricci curvature scalar is $\hat R$. 
 We use units $G = \hbar = c = 1$ and the  indices 
$\mu, \nu, ...$ take the values 0, 1, 2, and 3 while 
latin indices $a, b, ...$ will take the values 0 and 1. 
Integration over the angular variables lead to an overall 
factor of $4\pi$ from eq.  (\ref{eq:jap1}). 

{\rm  En route} towards our reduced model  we further 
take the following steps. To begin with, 
our 4-dimesional {\em spherically symmetric}
 metric is written as 
\begin{equation}
ds^2 =  h_{ab} dx^a dx^b + \phi^2 ( d\theta^2 + \sin^2\theta d\varphi^2 ),
\label{eq:jap2}
\end{equation}
together with the ADM decomposition 
\begin{equation}
h_{ab} = \left(
\begin{array}{cc}
-\alpha^2 + \frac{\beta^2}{\gamma} & \beta \\
\beta & \gamma
\end{array}\right) ~,~ 
h^{ab} = \left( 
\begin{array}{cc}
-\frac{1}{\alpha^2} & \frac{\beta}{\alpha^2\gamma} \\
\frac{\beta}{\alpha^2 \gamma} & 
\frac{1}{\alpha} - \frac{\beta^2}{\alpha^2\gamma^2}
\end{array}
\right),
\label{eq:jap3}
\end{equation}
where $\alpha, \beta, \gamma, \phi$ are functions 
of $(x^0, x^1) = (\tau, r)$ which will be defined later 
(see eq. (\ref{eq:vad2})). Hence ${\rm det} (h_{ab}) = - \alpha^2 \gamma$. 
We also take $\hat \psi = \psi (\tau, r)$ and 
\begin{eqnarray}
F_{ab} & = & \varepsilon_{ab} \sqrt{ -h} E; ~E = ( - h)^{-1/2} 
(\dot A_1 - A_0^\prime),  \label{eq:jap4} \\
\hat D_a \hat \psi & =  & \partial_a \psi + ieA_a\psi; 
\hat D_2 \hat \psi = \hat D_3 \hat \psi = 0.
\label{eq:jap5} 
\end{eqnarray}
Furthermore, we have that the trace of the extrinsic curvature is 
 written  as  
\begin{equation}
K =  \frac{\dot{\gamma}}{2\alpha\gamma} -  
\frac{\beta^\prime}{\alpha\gamma} + 
\frac{\beta\gamma^\prime}{2 \alpha \gamma^2},
\label{eq:jap7} 
\end{equation}
and which is present in the gravational part 
of the action (\ref{eq:jap1}) as 
(cf. ref. \cite{thomhaji,haji} for details)  
\begin{equation}
\frac{1}{2} \sqrt{- h} \phi^2 R = 
-\alpha\sqrt{\gamma} K \left(n^a\partial_a \phi^2\right) 
+ \frac{2\alpha^\prime \phi \phi^\prime}{\sqrt{\gamma}} 
+ {\rm total ~derivatives}.
\label{eq:jap8}
\end{equation}
Notice   we employ $``\cdot'' \equiv \frac{\partial}{\partial \tau}$ 
and $``\prime'' \equiv \frac{\partial}{\partial r}$ and 
$n^a$ is the normal unit to the $x^0 = {\rm constant}$ surfaces:
\begin{equation}
n^a = \left(\frac{1}{\alpha}, -\frac{\beta}{\alpha\gamma}
\right).
\label{eq:jap9}
\end{equation}

The next {\em significant}  step consists in  choosing  the 
following 
coordinate gauge: 
\begin{equation}
\alpha = \frac{1}{\sqrt{\gamma}} \Leftrightarrow \sqrt {- h} =1 
\label{eq:jap10}
\end{equation}
(see ref. \cite{pol} and in particular \cite{ryan} for a detailed 
explanation). Hence, we retrieve  the reduced 
Lagrangean\footnote{The Lagrangean (\ref{eq:jap11}) contains 
both $\alpha$ and $\gamma$ although the relation (\ref{eq:jap10}) 
states that there is a certain relationship between them. 
In fact, the reason for the particular form of (\ref{eq:jap11}) 
is {\em only} t simplify the canonical procedure to 
derive the Hamiltonian $H = \alpha {\cal H}_0 + 
\beta {\cal H}_1 + A_0 {\cal H}_2$ (see eq. (\ref{eq:jap12a}) 
-- (\ref{eq:jap13c}) .}
\begin{eqnarray}
{\cal L} & = & \left\{1 + 
\left[ - \frac{\dot{\phi}^2}{\alpha^2} + 
2 \beta\dot{\phi}\phi^\prime + \left(\frac{1}{\gamma} - 
\beta^2\right)(\phi^\prime)^2 \right] \right. \nonumber \\
& - & \left. [\dot{\gamma} - 2 \alpha^{-1} 
(\alpha\beta)^{\prime}] (\phi\dot{\phi} - \alpha^2\beta \phi 
\phi^\prime + \alpha\alpha^\prime \phi\phi^\prime\right\} \nonumber \\
& + & \frac{\phi^2}{\alpha^2} (\dot{\psi} + ieA_0\psi)
(\dot{\psi}^{\dagger} - ieA_0 \psi^\dagger) 
- \left(\frac{1}{\gamma} - \beta^2\right) \phi^2 
(\psi^\prime + ieA_1 \psi)(\psi^{\prime\dagger} - ieA_1 \psi^\dagger)
\nonumber \\
& - & \beta \phi^2 \left[
(\psi^\prime + ieA_1\psi) (\dot{\psi}^\dagger - ieA_0 \psi^\dagger) 
+ 
(\dot{\psi} + ieA_0 \psi \psi) (\psi^{\prime\dagger} - 
ieA_1 \psi^\dagger)\right] 
+ \frac{1}{2}\phi^2 \left(\dot{A}_1 - A_0^\prime\right)^2,
\label{eq:jap11}
\end{eqnarray}
from which we obtain the canonical momenta 
\begin{eqnarray}
\pi_\gamma & = & 
-\frac{1}{2}(\phi\dot{\phi} - \alpha^2 \beta \phi \phi^\prime), 
\label{eq:jap12a} \\
\pi_\phi & = & - \frac{\dot{\phi}}{\alpha^2} + 
\beta \phi^\prime - \frac{1}{2} 
(\dot{\gamma} - 2 \alpha^{-1} \alpha^\prime \beta - 
2 \beta^\prime) \phi 
\label{eq:jap12b} \\
\pi_\psi & = & 
\frac{\phi^2}{\alpha^2} (\dot{\psi}^{\dagger} - 
ieA_0 \psi^\dagger) - \beta \phi^2 (\psi^{\dagger\prime} - ieA_1 
\psi^\dagger) ,
\label{eq:jap12c} \\
\pi_{\psi^\dagger} & = & 
\frac{\phi^2}{\alpha^2} (\dot{\psi} + 
ieA_0 \psi) - \beta \phi^2 (\psi^{\prime} + ieA_1 \psi) ,
\label{eq:jap12d} \\
\pi_{A_1} & = & \phi^2 (\dot{A}_1 - A^{\prime}_0) 
\label{eq:jap12e} \\
\pi_\alpha & = & \pi_\beta = \pi_{A_0} = 0.
\label{eq:jap12f}
\end{eqnarray}
Hence,  the constraints can be written as 
\begin{eqnarray}
{\cal H}_0 & = & - 2 \phi^{-1} \sqrt{\gamma} \pi_\gamma \pi_\phi 
+ 2 \phi^{-2} \gamma^{3/2} \pi_\gamma^2 - 
\frac{1}{2} [\sqrt{\gamma} - \frac{\phi^{\prime 2}}{\sqrt{\gamma}} - 
2 \alpha^\prime \phi \phi^\prime] \nonumber \\
& + & \frac{\pi_\psi\pi_{\psi^\dagger}}{\sqrt{\gamma} \phi^2} 
+ \frac{\phi^2}{\sqrt{\gamma}}(\psi^\prime + 
ieA_1\psi) (\psi^{\prime\dagger} - ieA_1 \psi^\dagger) 
+ \frac{\sqrt{\gamma}}{2 \phi^2} \pi_{A_1}^2,
\label{eq:jap13a} \\
{\cal H}_1 & = & \frac{1}{\gamma}
\phi^{\prime}\pi_\phi - 2 \pi_\gamma^\prime \gamma^\prime 
+ \frac{1}{\gamma} (\pi_\psi \psi^\prime + ieA_1 \pi_\psi \psi 
+ \pi_{\psi^\dagger} \psi^{\dagger} -ieA_1 \pi_{\psi^\dagger}
\psi^\dagger), 
\label{eq:jap13b}  \\
{\cal H}_2 & = & -ie(\pi\psi_\psi - \psi^\dagger \pi_{\psi^\dagger}) 
- \pi_{A_1}^\prime.
\label{eq:jap13c}
\end{eqnarray}

Regarding the quantization of this system, we will analyse it 
in the vicinity of a  suitable  apparent 
horizon \cite{haji}. In this case, $\alpha, \beta, \gamma$ are all finite 
and non-zero. Self-consistency conditions will be found 
and/or imposed on the ``gravitational'' constraints 
(\ref{eq:jap13a}), (\ref{eq:jap13b}) and the gauge 
constraint (\ref{eq:jap13c}). This is the subject of the 
following section.

\section {Description of a RN Black-Hole in the Apparent 
Horizon}

\indent

In this section, we will express 
the constraints (\ref{eq:jap13a}),  (\ref{eq:jap13b}) and 
(\ref{eq:jap13c})  in terms of dynamical quantities defined at  an 
apparent horizon of the RN black-hole, 
which is defined by the condition \cite{haji}
\begin{equation}
h^{ab}(\partial_a \phi)(\partial_b \phi) = 0 \Leftrightarrow 
\dot{\phi}(\dot{\phi} - \phi^\prime) = 0.
\label{eq:p31}
\end{equation}
Moreover, we will follow M. Pollock's approach \cite{pol}, 
which involves some differences with respect to the method present 
in ref. \cite{tomi, odat, odasolo}.

The first element of our (and Pollock's)  approach was the introduction 
of the coordinate gauge choice (\ref{eq:jap10}) 
in the action (\ref{eq:jap11}), as 
described in the previous section. This seems to introduce 
differences in the 
Hamiltonian structure. In fact, we have (and in ref. \cite{pol} 
as well) a term 
$\alpha^\prime \phi \phi^\prime$ in ${\cal H}_0$ (see eq. 
[\ref{eq:jap13a}) above and then the expression for $W$ in pg. 1180 
of ref. \cite{thomhaji} and subsequently eq. (8) in that ref. as well]
after having employed 
 $\frac{1}{\sqrt{\gamma}} = \alpha$. In ref. \cite{odat, odasolo} 
an integration by parts is employed to get $\frac{\alpha^\prime 
\phi \phi^\prime}{\sqrt{\gamma}} \rightarrow - \alpha \left[ 
\frac{\phi \phi^\prime}{\sqrt{\gamma}}\right]^\prime$ without yet using 
a gauge condition as (\ref{eq:jap10}). 

In order to obtain a satisfactory 
description of an evaporating RN black 
hole on the apparent horizon, it is more 
convenient to use a RN - Vaidya metric \cite{RNvad, fayos}. This 
 has the general form
\begin{equation}
ds^{2}  = -\left(1 - \frac{M(v)}{r} 
+ \frac{Q^{2}}{r^{2}}\right)dv^{2} 
+ 2dvdr + \phi^{2}d\Omega_{2}^{2}.
\label{eq:vad1}
\end{equation}
Here $v \equiv  \tau + r = t + r^*$ is the 
advanced null Eddington-Finkelstein coordinate, 
$r^*$ is the corresponding ``tortoise" 
coordinate and the relationship between 
the time $\tau$ and the time coordinate 
$t$ for the standard  RN metric\footnote{Standard RN metric: $ds^2 = 
-\left(1 - \frac{2M}{r} + \frac{Q^2}{r^2}\right) dt^2 
+ \left( 1 - \frac{2M}{r} + \frac{Q^2}{r^2}
\right)^{-1} dr^2 + r^2 d\Omega_2^2$.} is 
$\tau = t - r + r^*$. In addition, $d\Omega^2_{2}$ 
represents the area element of the two-sphere: 
$d\theta ^2 + \sin ^2\theta d\varphi ^2$.
Notice that we are treating $Q$ as a constant.

At  this point  we will further include and adapt more 
elements from the method presented in \cite{pol} to our  RN 
black hole case. First, we use the coordinates $(\tau, r, \theta, \varphi)$. 
In terms of the coordinates 
$(\tau, r, \theta, \varphi )$ the 
Vaidya RN metric becames
\begin{equation}
ds^2 = -\left(1 - \frac{2M}{r} 
+ \frac{Q^2}{r^2}\right)d\tau ^2 
+ \left(\frac{4M}{r} 
- \frac{2Q^2}{r^2}\right)d\tau dr 
+ \left(1 + \frac{2M}{r} - 
\frac{Q^2}{r^2}\right)d r^2 + 
\phi^2d\Omega _2^2,
\label{eq:vad2}
\end{equation}
where the quantities $\alpha, \beta, \gamma$ 
take the form
\begin{equation}
\alpha = \left(1 + \frac{2M}{r} - 
\frac{Q^2}{r^2}\right)^{- 1/2}, ~
\beta = \frac{2M}{r} - \frac{Q^2}{r^2}, ~
\gamma = 1 + \frac{2M}{r} - 
\frac{Q^2}{r^2}.
\label{eq:qtvad2}
\end{equation}
The RN black-hole has two apparent horizons, namely at 
\begin{equation}
r_{\pm}(v)  = M (v) \pm \sqrt{M^2(v) - Q^2},
\label{eq:apahoriz}
\end{equation}
and we will henceforth restrict ourselves to 
the case of $r_+$. 
Secondly, in similarity witht the Schwarzschild case 
we also  take $M \simeq M(\tau)$ in the vicinity of the apparent horizon
(see ref. \cite{pol, RNvad, fayos, hisc} for details). 

After some lenghty calculations and using 
the approximations 
$\phi \simeq r, M\simeq M(v)  \sim M(\tau)$ 
we obtain that at the apparent horizon $r_+$ 
the following quantitites can be {\em exactly} written as 
\begin{eqnarray}
\alpha & = &\frac{1}{\sqrt{2}}, ~
\beta = 1, ~\gamma =2, \label{eq:qtvad3} \\
h_{00}& = & 0, ~h_{01} = 1, ~h_{11} = 2, \label{eq:qtvad3aa} \\
h^{00} & =  &-2, ~h^{01} = 1, ~h^{11} = 0, \label{eq:qtvad3a}\\
\beta ^\prime & = &\frac{2}{\rho}\left (\frac{M}{\rho} 
- 1\right ), \label{eq:qtvad3new} \\
~ \alpha ^\prime & = &  
-\frac{1}{2\sqrt{2}\rho}\left (\frac{M}{\rho} 
- 1\right ), \label{eq:qtvad3b}\\
\dot{\gamma }& = & 2\frac{\dot M}{\rho}, \label{eq:qtvad3newa} \\
~\gamma ^\prime & = &  \frac{2}{\rho}\left(\frac{M}{\rho} 
- 1\right ),~ \label{eq:qtvad3c} \\
\pi^\prime_{\gamma }& = & - 
\frac{1}{4}\left(1 - \frac{M}{\rho}\right) 
+ \frac{1}{4}, ~\pi_{\gamma} = \frac{\rho}{4}, 
\label{eq:piphint} \\ 
\pi_{\phi}& = &  -\dot{M} - \frac{1}{2} + 
\frac{3}{2}\frac{M}{\rho}, 
\label{eq:piphitrue} \\
\pi_{A_{1}} & = &  \rho^{2}(\dot{\bar A_{1}} 
- \bar A_{0}^\prime ), \label{eq: qtvad3d} \\
\pi_{\psi}& = & 2\rho^2 [\dot{\bar \psi }^\dagger - 
\bar \psi^{\prime \dagger } - ie(\bar A_0 - \bar A_1)
\bar \psi ^\dagger ], \label{eq:qtvad3newbb} \\
\pi_{\psi ^\dagger} & =  &  2 \rho^2 [\dot {\bar \psi } - 
\bar \psi^{\prime } + ie(\bar A_0 - \bar A_1)\bar \psi ], 
\label{eq:qtvad3e}
\end{eqnarray}
which agree with the corresponding expressions in the case of 
$Q=0$ and $\rho = 2M$ (see ref. \cite{pol}).
A ``overline'' means the value of the variable 
taken at the vicinity of $r_+$. 
Employing (\ref{eq:qtvad3})-(\ref{eq:qtvad3e}), 
the constraint equations are then written 
as\footnote{Notice that first we ought to 
find the explicit expressions for  the variables $\alpha, 
\beta, \gamma, \phi, \psi, \psi^\dagger$, etc, and their 
time/spatial derivatives 
(see e.g., eq. (\ref{eq:jap12a})--(\ref{eq:jap12f}), 
(\ref{eq:qtvad2})). {\em Only ~afterwards} 
we calculate their value at the apparent horizon: see eq. 
(\ref{eq:qtvad3})--(\ref{eq:qtvad3e}). Subsequently, we  
can then employ these quantitites in eq. (\ref{eq:jap13a})--(\ref{eq:jap13c}) 
and retrieve eq. (\ref{eq:H0a})--(\ref{eq:H2a}).
as described  above.}
(droping the overline ``bar'' henceforth) 
\begin{eqnarray}
{\cal H}_0 & = & \frac{1}{2}\pi_\phi -
\frac{1}{4} + \frac{M}{4\rho} - 
\frac{1}{2\rho^2} \pi_{\psi}\pi_{\psi^\dagger } - 
\frac{1}{2\rho^2}\pi_{A_1}^2 
- \frac{\rho^2}{2}(\psi^\prime + 
ieA_1\psi) (\psi^{\prime \dagger } - 
ieA_1\psi^\dagger)
\label{eq:H0a}\\
{\cal H}_1 & = & \frac{1}{2}\pi _\phi + 
\frac{1}{4} - \frac{3M}{4\rho} + 
\frac{1}{2}[\pi _\psi (\psi ^\prime 
+ ieA_1\psi ) + 
\pi _{\psi ^\dagger }(\psi ^ {\prime \dagger } 
- ieA_1\psi ^\dagger)]
\label{eq:H1a}
\\
& = & \frac{1}{2}\pi _\phi + 
\frac{1}{4} - \frac{3M}{4\rho} + 
\frac{1}{2}(\pi _\psi \psi^\prime + 
\pi_{\psi^\dagger }\psi^{\prime \dagger}) - 
\frac{A_{1}}{2}({\cal H}_2 - 
\pi_{A_{1}}^\prime )]
\label{eq:H1aa}\\
{\cal H}_2 & = & -ie(\psi \pi _\psi - 
\psi^{\dagger}\pi_{\psi^\dagger}) - 
\pi _{A_1}^\prime,
\label{eq:H2a}
\end{eqnarray}
where we are employing $\rho \equiv r =  r_+ =  M + 
\sqrt{M^2 - Q^2}$.  Later on, we will use the approximation 
$\rho \sim  2M - 
\frac{Q^2}{2M}$. In the following 
and where appropriate, 
  we will  be 
replacing $\phi \simeq  r$ by the preceding 
expression $\rho (M)$ at the  apparent 
horizon $r_+$. Hence, $\dot {\phi } = 0$ 
and $\phi^\prime \neq 0$ and eq. (\ref{eq:p31}) 
is satisfied.

It is also worthy  to notice the folowing properties, regarding 
our model prior to expressing eq. (\ref{eq:jap13a}) and 
(\ref{eq:jap13b}) 
in quantitites evaluated at the apparent horizon $r_+$. 
The ``purely'' geometric terms
in eq. (\ref{eq:jap13a}),  (\ref{eq:jap13b}) 
(first, second, third and first, second, respectively)
are the same 
either in the Schwarzschild or  RN cases. Moreover, 
the values of $\alpha, \beta, \gamma$ at the apparent horizon 
are also the same 
either in the Schwarzschild or  RN cases. 
However,  the value of their spatial derivatives 
at $r_+$ is different (see above and ref. 
\cite{pol}): 
the presence of the charge $Q$ induces a different geometry 
and the rate of change along spatial geodesics is different from the 
Schwarschild black-hole. If $Q=0$, $\rho = 2M$ then 
 we get a proportionality 
between those  geometrical terms in ${\cal H}_0$ and ${\cal H}_1$ at the 
apparent horizon,  in {\em total} agreement with ref. \cite{pol}
(cf. the first, second and third terms in 
either eq. (\ref{eq:H0a}) and (\ref{eq:H1a})). Moreover, we should 
also 
stress that  the ``geometrical'' terms in 
 (\ref{eq:H0a}) and (\ref{eq:H1a}) 
are different from the corresponding ones in \cite{odat, odasolo}.
We will further discuss this aspect in section 5.

Another important characteristic 
of the method 
employed in ref. \cite{pol} 
(see also ref. \cite{pol, fayos, hisc}) is as follows. The RN-Vaidya 
metric requires (in $d = 4$ dimensions) a trace-free
matter source  energy momentum tensor, which in our case 
involves  
 complex scalar fields 
$\psi, \psi ^\dagger$ with the $A_\mu$ 
field. Hence,  we  obtain
\begin{eqnarray}
T_{matter}& = & - \frac{3}{2}(\dot{\psi }^\dagger \dot{\psi } 
+ e^2A_0^2\psi ^\dagger \psi + 
ieA_0\dot{\psi }^\dagger \psi - 
ieA_0\psi ^\dagger \dot {\psi})
\nonumber\\ 
& + & \frac{3}{4} (\dot{\psi }^\dagger \psi^\prime 
- e^2 A_0 A_1 \psi^\dagger \psi + 
ie A_1 \psi \dot{\psi}^\dagger
- ie A_0 \psi^\dagger \psi^\prime ),
\nonumber\\
& + & \frac{3}{4} (\dot{\psi }\psi^{\prime \dagger } 
- e^2 A_0 A_1 \psi^\dagger \psi + 
ieA_0 \psi {\psi }^{\prime \dagger }   
 - ie A_1  \psi^\dagger \dot \psi) = 0.
\label{eq:trace}
\end{eqnarray}
A simple possibility for a boundary condition 
extracted from (\ref{eq:trace}) is 
\begin{equation}
\dot{\psi}^\dagger(\dot{\psi} - \psi^\prime) 
+ \dot{\psi}(\dot{\psi}^\dagger - 
\psi^{\prime\dagger}) = 0, ~ A_0 = 0,
\label{eq:bc}
\end{equation}
which has important  similarities to what 
Pollock \cite{pol} introduced for the Schwarschild case.

Let us also 
introduce the following re-scaling 
which will allow us to monitor our case 
in close comparison with M. Pollock's \cite{pol} 
and also to address the compatibility 
of these constraints:
\begin{eqnarray}
\psi & = & \sqrt{2\pi }(\psi_1 + i\chi )
\label{eq:redef1a} \\
\pi_{\psi } & \rightarrow & \frac{1}{\sqrt{8\pi }} \pi_\psi,
\label{eq:redef1b}
\end{eqnarray}
and similarly for their conjugates. 
We then get the approximate expressions  at $r_+$: 
\begin{eqnarray}
{\cal H}_0 & = & \frac{1}{2}\pi_\phi - 
\frac{1}{8} + \frac{Q^2}{32M^2} - 
\frac{1}{16\pi }\frac{1}{\rho^2}(\pi ^2_{\chi } 
+ \pi_{\psi_1}^2) - \rho^2\pi (\psi_1^{\prime^2} +  
 \chi^{\prime 2}) \nonumber \\
& - &  \frac{1}{2\rho^2}\pi _{A_1}^2 - 
\frac{\rho^2}{2}[2\pi e^2 A_1^2 (\psi_1^2 
+ \chi^2) ] - 2 \pi e \rho^2 A_1 (\psi_1\chi^\prime - 
\chi \psi_1^\prime )]
\label{eq:H0b}\\
{\cal H}_1 & = &\frac{1}{2}\pi_\phi - 
\frac{1}{8} - \frac{3Q^2}{32M^2} + 
\frac{1}{2}[\pi_{\psi_1} \psi_1^\prime + 
\pi_{\chi} \chi^\prime] + \frac{e A_1}{4}
(\pi_{\psi_1} \chi  - \pi_\chi \psi_1)  
\label{eq:H1b}
\end{eqnarray}

Concerning the compatibility of the ${\cal H}_0$ and 
${\cal H}_1$ constraints, the following points are now in order. 
As we mentioned earlier, it  
is worthwhile to notice that as far as the 
``purely geometrical" terms in 
(\ref{eq:jap13a}) and (\ref{eq:jap13b}) are concerned, 
an {\em exact} proportionality    
${\cal H}_0 = C{\cal H}_1$ can only be achieved  
if we choose $\rho = 2M$ ($Q=0$), where $C$ is some constant.
In our case, however, we can only have an {\em approximatte}  
proportionality for these terms and to this end we will require 
a small value for $\mid Q\mid$ 
in comparison with $M$. Then, compatibility 
for the RN ``geometrical'' terms above 
(and let us emphasize, {\em within the method} of 
ref. \cite{pol} that we are {\em testing} in the RN case) 
could  be satisfactory up to 
terms of order $\frac{Q^2}{M^2}$, which 
will be considered as a physical perturbation. 
We will further 
 take $\chi$ as a small perturbation and hence 
will consider  any electromagnetic related 
terms as with  a {\em very} 
 small magnitude. 

Moreover, compatibility between the fourth and fifth terms in  
(\ref{eq:H0b}) and the fourth term in (\ref{eq:H1b}) 
requires that (see ref. \cite{pol} for a comparison) 
\begin{equation}
\pi_{\psi _1} = -4\pi \rho^2\psi_1^\prime, ~ 
\pi_\chi = - 4 \pi \rho^2 \chi^\prime.
\label{eq:consis1}
\end{equation}
Quite  interestingly, (\ref{eq:consis1}) also constitutes  
the precise  compatiblity condition that is necessary between 
the eight term in (\ref{eq:H0b}) and the last terms in 
(\ref{eq:H1b}).

Another  option at this point is to require   $A_1$ 
to be small,  and then take the terms with $e \rho A_1
 \psi_1, e \rho A_1 \chi , 
e  A_1 \psi_1, e  A_1 \chi, 
\frac{\pi_{A_1}}{\rho}$ to  be 
negligible. Subsequently,  we could proceed with solving ${\cal H}_0 
\simeq 
{\cal H}_1 = 0$.

With respect to the remaining constraint in 
eq. (\ref{eq:H2a}), notice that  the fifth term in eq. (\ref{eq:H1b}) 
can be  written basically as $A_1 ({\cal H}_2 - \pi^\prime_{A_1})$, 
after having  employed eq. (\ref{eq:H1aa}) and  (\ref{eq:H2a}).
Let us now choose to assume that our RN black hole case is such 
that allows to have\footnote{This is indeed 
a {\em very particular} choice but our physical results 
(see eq. (\ref{eq:wdw3}), (\ref{eq:solt1})). In fact, we could 
alternatively take the other restricting case of 
$\psi = \psi_1 + i\chi \rightarrow \psi = \psi_1$ with no 
gauge constraint (see ref. \cite{odat, odasolo}). We will then 
obtain solutions like (\ref{eq:solt1}) as a generalization 
of the ones in \cite{pol} but {\em without} the functional $F$. The
purpose of our choice is just to find a very particular situation of a 
case where $\chi$ and $A_1$ could be present in 
$\Psi_{{\rm RN}}$, even if in a very limited case.}
 ${\cal H}_1 \equiv 
H_1 \oplus 
A_1 ({\cal H}_2 - \pi^\prime_{A_1}) = 0$,
with $H_1 \gg A_1 ({\cal H}_2 - \pi^\prime_{A_1}) 
\simeq  0$. Note that in our choice of approximation we are taking 
terms like  $e A_1 \chi$ as very small when 
compared with others as, e.g., $\pi_{\chi}^2$ in 
(\ref{eq:H0b}),   and so  the 
fifth term in eq. (\ref{eq:H1b}) is being  taken as substacially 
less influential than the others
Remember now  that
by construction, we have 
${\cal H}_2 = 0$. 
If we take 
$H_1 \gg A_1 ({\cal H}_2 - \pi^\prime_{A_1}) 
\simeq  0$ then  ${\cal H}_2 = 0$ implies $\pi_{A_1}^\prime = 0$. 
So, within this last restriction we further take 
$H_1 = 0$ (as just defined) and also 
${\cal H}_2 = \pi^\prime_{A_1} = 0$. 

Let us finally also mention that eq. (\ref{eq:consis1}) implies from 
eq. (\ref{eq:trace}) that at the apparent horizon 
we may  take the conditions $A_0 = 0$ together   with 
$\dot \psi = \dot \psi^{\dagger} = 0$ (see ref. \cite{pol} 
for the 
Schwarzschild case). 
However, we are only  imposing these conditions and restrictions 
 {\em after} varying the action and obtaining the constraint equations. 
We further take $A_1 = A_1 (r)$.

\section {Quantum States from the Wheeler-DeWitt 
Equation}

\indent

As described above, the restrictions and boundary 
conditions introduced in \cite{pol} and adapted here to the 
RN black hole case
lead to the approximate 
Wheeler-DeWitt equation 
(up to terms in $Q^2/M^2$ and neglecting terms 
$A_1$-related  which are taken of small magnitude) 
  at the 
apparent horizon $r_+$, 
\begin{equation}
\pi _\phi \simeq  \frac{\pi ^2_{\psi _1}}{4\pi \rho^2(M)} 
+ \frac{\pi ^2_{\chi }}{4\pi \rho^2(M)} + 
\frac{1}{4}.
\label{eq:wdw1}
\end{equation}
Quantization proceeds via the operator 
replacements
\begin{eqnarray}
\pi_{\phi }& \rightarrow & - 
i\frac{\partial }{\partial \phi } \simeq  
-i\frac{2M^2}{4M^2 + Q^2}\frac{\partial }{\partial M},
\label{eq:mom1}\\
\pi_{\psi_1}& \rightarrow & - 
i\frac{\partial }{\partial \psi _1},
\label{eq:mom2}\\
\pi_{\chi }& \rightarrow & - 
i\frac{\partial}{\partial \chi },
\label{eq:mom3} 
\end{eqnarray}
which yelds the Wheeler-DeWitt equation 
for the wave function $\Psi$,
\begin{equation}
- i 
\frac{2M^2}{4M^2 + Q^2} \frac{\partial \Psi}{\partial M}= - 
\frac{1}{16\pi M^2 + \pi Q^4/M^2 - 8\pi Q^2} 
\left[\frac{\partial^2 \Psi }{\partial \psi_1^2} 
+ \frac{\partial^2 \Psi }{\partial \chi^2} \right ] 
+ \frac{1}{4}\Psi.
\label{eq:wdw3}
\end{equation}
Eq. (\ref{eq:wdw3})  has 
a Schr\"odinger-like form and 
introducing 
$\Psi = \hat \Psi (M)\tilde \Psi (\psi_1, \chi ) \zeta(A_1)$ 
we get the equations
\begin{eqnarray}
\frac{\partial \hat \Psi }{\partial M}& = & \left [ i\frac{8\pi M^4 - 
2\pi Q^4 - 4\pi M^2Q^2 + \pi Q^6/M^2}
{32\pi M^4 + 2\pi Q^4 - 16\pi Q^2M^2} + 
ik^2\frac{2M^2 + Q^2}{32\pi M^4 + 2\pi Q^4 - 
16\pi Q^2M^2}\right ]\hat \Psi
\label{eq:wdw4a}\\
\frac{\partial^2 \tilde \Psi }{\partial \psi_1^2}& 
+ & \frac{\partial^2\tilde \Psi }{\partial \chi^2}
= - k^2\tilde \Psi,
\label{eq:wdw4b}
\end{eqnarray}
whose solutions are 
\begin{equation}
\Psi_{{\rm RN}} [
M, Q; \psi_1, \chi; k]  = \Psi^0_{{\rm RN}} 
e^{i\left [\frac{1}{4}\left (- \frac{Q}{M} 
+ 2M\right ) + k^2\frac{M}{Q^2 - M^2}
\pm 2\sqrt {\pi }k(\psi _1 + \chi ) \right]}
\zeta (A_1),
\label{eq:solt1}
\end{equation}
where $k^2$ is a separation constant and 
$\Psi^0_{{\rm RN}}$ an integration constant.
There seems to be no adequate procedure to fix the 
parameter $k$ without introducing new physics. However, the Schr\"odinger 
form of eq. (\ref{eq:wdw3}) implies the possibility of positive 
semi-definite probability densities $\Psi \Psi^*$. This may suggest that a 
black hole could evaporate (see below) without violation of unitarity.

 In the very particular case of 
${\cal H}_2 \simeq  \pi_{A_1}^\prime \simeq  0$ then we could take 
 $\pi_{A_1} \rightarrow -i \frac{\partial}{\partial A_1}$
and get the  equation 
\begin{equation}
\frac{d}{d ~r} \left( 
\frac{\partial \zeta[A_1(r)]}{ \partial A_1 (r)}\right)  = 0, 
\end{equation}
which allows to add 
 (cf. ref. \cite{ki})
\begin{equation}
\Psi_{{\rm RN}} [
M, Q; \psi_1, \chi; k]  = e^{i\left [\frac{1}{4}\left (- \frac{Q}{M} 
+ 2M\right ) + k^2\frac{M}{Q^2 - M^2}
\pm 2\sqrt {\pi }k(\psi _1 + \chi ) \right]}
F \left(\int_{-\infty}^{+\infty} d r A_1  \right), 
\label{eq:solt1a}
\end{equation}
where $F$ is an arbitrary function. 
The  point to notice is that 
we now have explicit solutions concerning the dependence in 
$M$ and $Q$ of (\ref{eq:solt1}).

It is interesting to notice the following as well. For the Schwarzschild case 
($Q=0)$ eq. (\ref{eq:solt1a}) implies that near to $M=0$ the wave 
function will oscilate with infinite frequency. If $\dot{M} < 0$,  this would 
represent the quantum mechanical behaviour of the black hole 
near the end point of its evaporation. In the RN case, the 
rapid oscillations 
will occur again for $M=0$ but also when 
$M \sim Q$. I.e., near extremality and when the black hole 
mass evaporation 
can eventually  stop. Hence, the presence of $Q$ in $\Psi_{{\rm RN}}$ 
allowed us to identify some known   physical situations of the 
RN bkack hole. 
Moreover, when $Q \neq 0$ we have more and different values for $M$ which 
lead to $\hat \Psi \sim {\rm constant}$.

As far as the mass-charge ratio  for the RN 
black hole is concerned, we will use eq. (\ref{eq:piphitrue}) for 
$\pi _\phi$, i.e.,
\begin{equation}
-i\frac{\partial \Psi_{{\rm RN}}}{\partial \phi} = 
\left[ - \dot{M} - \frac{1}{2} + 
\frac{3M}{4M^2 - 2 Q^2/M}\right] \Psi_{{\rm RN}}. 
\label{eq:mrateo}
\end{equation}
From eq. (\ref{eq:solt1}) and eq. (\ref{eq:mrateo}) and 
get the equation\footnote{It should be emphasized that at this point
we are taking $M$ as an expectation  value (average), i.e., 
semiclassical value, over the sates $\Psi_{{\rm RN}}$.}
\begin{equation}
\dot{M} + 
\frac{1}{4M^2} \frac{a[k^2; Q]}{d(M)} + 
\frac{1}{4M^4} \frac{b[k^2; Q]}{d(M)}
+ \frac{c[k^2; Q]}{d(M) M^6}
= 0,
\label{eq:newlast1}
\end{equation}
where
\begin{eqnarray}
a[k^2; Q] & = & k^2 - 5Q^2 + Q, 
\label{eq:newlast2a} \\
b[k^2; Q] & = & k^2 Q^2 - 3Q^4, 
\label{eq:newlast2b} \\
c[k^2; Q] & = & \frac{k^2 Q^4}{8}, 
\label{eq:newlast2c} \\
d(M) & = & 1 + \frac{Q^2}{2M^2}.
\label{eq:newlast2d} 
\end{eqnarray}
An  integration of (\ref{eq:newlast1}) leads to the approximate 
result
\begin{equation}
M = \left[ M_0^3 - \frac{3}{2}(k^2 - 5 Q^2 + Q)\right]^{1/3} 
(t - t_0)^{1/3} .
\label{eq:mrate2}
\end{equation}
We can now identify several physical cases of interest for the RN 
black hole, according if $a, b, c, d$ are either positive, zero or negative.

For the case of $Q=0$ \cite{pol} (Schwarzschild),  it is the separation 
constant $k^2$ that determines if the black hole is evaporating and 
decreasing its mass ($k^2 >0$), or increasing its mass 
($k^2 < 0 \Leftrightarrow k$ imaginary). In the present RN 
black hole case, the presence of the charge $Q$ introduces significative 
changes. In fact, notice that $d>0$ and $c>0$ (if $k^2<0$) but 
$a\leq 0$ when $Q \geq \frac{1 + \sqrt{1 + 20 k^2}}{10}$ or 
$Q \leq \frac{1 - \sqrt{1 + 20 k^2}}{10}$ (if $k^2 <0$), 
and $ b \leq 0$ when $Q \leq - \sqrt{\frac{k^2}{3}}$ or 
when $Q \geq  \sqrt{\frac{k^2}{3}}$ (if $k^2>0$). 
If $k^2 < 0$, then $b\leq 0$ ($b=0 \Leftrightarrow Q=0$)
and $a$ can only be positive if $1 + 20k^2 > 0$. When 
$a>0, b>0$, then the RN black hole mass decreases (if $k^2  >0$). 
In this case, we have an evaporating RN black hole with probability 
density $\varrho \equiv \Psi_{{\rm RN}} \Psi^*_{{\rm RN}} = 
(\Psi_{{\rm RN}}^0)^2$, independent of $M, Q, \psi_1$ and $\chi$, 
which expresses the existence of the hole.

A $\dot{M} > 0$ stage can be obtained from eq. (\ref{eq:mrateo}), 
with $k^2 >0$ but $a<0, b<0$ and $c>0, d>0$. If $Q=0$, this possibility is 
absent. When $Q\neq 0$, $k^2 < 0$, then $c<0, b<0, a<0$, if 
$1 + 20 k^2 > 0$ has real solutions. The latter situation leads to 
$\varrho \equiv \Psi_{{\rm RN}} \Psi^*_{{\rm RN}} = 
(\Psi_{{\rm RN}}^0)^2 e^{-2\mid k \mid \mid (\psi + \chi)\mid}$,
assuming a well behaved solution $\Psi_{{\rm RN}}$. 
For the former, we have again that 
$\varrho \equiv \Psi_{{\rm RN}} \Psi^*_{{\rm RN}} = 
(\Psi_{{\rm RN}}^0)^2$. 

\section {Conclusions and Discussion}

\indent

The main purpose of this letter was to contribute 
with an additional and different  perspective on the 
quantization of RN black holes. 
The physics of black holes is indeed a fascinating 
subject and we trust our original 
results may add new information regarding  the many approaches 
to the quantization of charged black holes.

Our canonical quantized 
RN black hole model has  particular characteristics as far as others 
(see ref. \cite{odat,odasolo})  are concerned. Among  these 
 is the fact that  we extended the 
framework introduced in   ref. \cite{pol} to a 
RN geometry in the presence of a complex 
scalar fields and a vector field $A_\mu$. The method of 
A. Tomimatsu \cite{tomi} was instead 
used in  ref. \cite{odat,odasolo} for  
the case of the Schwarzschild and RN 
black holes.
Moreover, we  employed the Hamiltonian formulation  
present in \cite{thomhaji,haji}, which was also 
used in ref. \cite{odat,odasolo} and with which this 
work could be compared. 

One of the main differences between this letter and 
ref. \cite{ odat, odasolo} corresponds to the expressions 
for the ${\cal H}_0$ and ${\cal H}_1$ constraints 
(\ref{eq:H0a}) and (\ref{eq:H1a}) {\em at} the apparent 
horizon $r_+$. Our results were thoroughly checked and confirmed, hence 
the reason for  these  differences could be identified with 
the restrictions that 
M. Pollock's approach \cite{pol} induces. Namely, the 
coordinate gauge choices as (\ref{eq:jap10}), 
subsequent restrictions (\ref{eq:trace}), 
(\ref{eq:bc}) with $Q = {\rm constant}$, as well as 
choosing the $(\tau, r, \theta, \varphi)$ frame. 
In fact, we mentioned in the begining of section 3 that one of the main 
elements in the approach introduced in ref. \cite{pol} (and used here) 
is the introduction 
of the gauge choice (\ref{eq:jap10}) in the action (\ref{eq:jap11}).
 This seemed to introduce differences in the 
Hamiltonian structure. In fact, we have (and in ref. \cite{pol} 
as well) a term 
$\alpha^\prime \phi \phi^\prime$ in ${\cal H}_0$ [see eq. 
(\ref{eq:jap13a})  and then the expression for $W$ in pg. 1180 
of ref. \cite{thomhaji} and subsequently eq. (8) in that ref. as well] 
after employing 
 $\frac{1}{\sqrt{\gamma}} = \alpha$. In ref. \cite{odat, odasolo} 
an integration by parts is employed to get $\frac{\alpha^\prime 
\phi \phi^\prime}{\sqrt{\gamma}} \rightarrow - \alpha \left[ 
\frac{\phi \phi^\prime}{\sqrt{\gamma}}\right]^\prime$.

Nevertheless, let us emphasize again that Pollock's method 
\cite{pol} and Tomimatsu-Hosoya-Oda's \cite{tomi, odat, odasolo} 
give equivalent descriptions in the Schwarzschild case. Only for the 
RN case there seems to be differences. 
Thus, our results 
also constitute a test on the method expressed in \cite{pol} 
and how it can be suited to black hole cases other than the 
Schwarzschild space-time. Moreover, it can 
 further inform on the scope of validity of different 
canonical formulations which seem to be equivalent 
in the Schwarzschild case \cite{pol, tomi},  but 
apparently lead to some differences in the RN case.

Let us also mention that we used 
 $\psi = \psi _1 + i\chi$, where  
$\chi$ could  be  interpreted as 
less physically relevant than $\psi_1$ 
from   conditions we ought to impose. In addition, 
several other restrictions and  approximations 
had to be introduced within the application of ref. 
\cite{pol} method for the RN case
as  we 
explained in section 3. Only by doing so we could 
get an approximated consistency and 
deal satisfactorily with 
the quantization at  the apparent 
horizon, but needing  the 
charge $\mid Q\mid$ to be  smaller when 
compared with the mass $M$. This poses obvious 
limits in the validity of our comments for $M\rightarrow |Q|$ but 
which are still of some qualitative interest.

As a consequence, we obtain quantum 
states from a  Wheeler-DeWitt 
(Schr\"odinger-like) equation. This one is  
present in eq. (\ref{eq:wdw3}) and  solutions are found 
in eq. (\ref{eq:solt1}). 
The case $(i)$ $Q=0$ corresponds to the Schwarzschild case and the 
corresponding 
wave function
implies a period of infinitely rapid  of oscilations near $M\sim 0$. 
In the 
RN case, this also occurs  near $M \sim 0$ but  near 
$M \sim \mid Q \mid$ as well, i.e., when the 
RN black hole is approaching extremality. Notice that only in 
this situation the RN black hole  has 
 supersymmetric  properties.

Can our model have an interpretation with 
particles coming out from the hole? 
We  could 
interpretate our analysis in such a way 
that it predicts a flow of complex 
scalar particles coming out from the hole. 
This conclusion seems natural on 
grounds
that we  are assuming a spherically symmetric 
electromagnetic field. Since
the physical photons correspond to 
transverse wave modes, it seems to me that 
an
assumption of a spherical symmetric electromagnetic field excludes a
possibility of photons coming out from the hole. A generalization of this
 to include some spherical asymmetry to our electromagnetic field might
perhaps merit further study in forthcoming publications.

Finally, we found that the RN mass-charge ratio permited  a 
$\dot{M} > 0$ and $\dot{M} < 0$ stages. The latter 
can be associated with usual black hole mass evaporation
while the former may require further analysis. The stage with 
$\dot{M} >0$ could suggest a physical effect in the terms of mass inflation 
\cite{minf} or an (in)direct  consequence of it. However, our approach 
(having followed ref. \cite{pol} method's) seems to be  of limited validity
and these precise claims must be taken with some caution. 

Nevertheless,  our results 
do bring additional and complementary information 
regarding other recent research \cite{odat, odasolo}.
In particular, by testing the approach of \cite{pol} 
in RN black holes and identifying its limits of application, 
the method of Tomimatsu-Hosoya-Oda seems to embrace much 
more physical situations of gravitational collapse.  
There is still 
the need for further investigations in the topic of
 black hole quantization, which one hopes may provide some 
interesting insights on the issue of 
canonical quantization of black holes 
but in a N = 2 supergravity 
theory \cite{moniz1}..

\vspace{1cm}

{\bf Acknowledgments}

This work was supported by the JNICT/PRAXIS XXI Fellowship 
BPD/6095/95. The author is  grateful to M. Cav\`aglia, 
and J. Louko   for interesting conversations and 
 discussions, as well as to 
 J. M\"akel\"a  for further important comments and suggestions.
 He also wishes to thank I. Oda 
for letting him know of his recent work and for extensive correspondence. 
Tlechsoha typed part of this report for which the author is most 
grateful.

\end{document}